\documentclass[12pt]{article}
\usepackage{graphicx}

\newcommand\pubdate{\today}

\def\Title#1{\begin{center} {\Large #1 } \end{center}}
\def\Author#1{\begin{center}{ \sc #1} \end{center}}
\def\Address#1{\begin{center}{ \it #1} \end{center}}

\newcommand\pubblock{\rightline{\begin{tabular}{l}  \\ 
         \pubdate  \end{tabular}}}
\newenvironment{Abstract}{\begin{quotation}  }{\end{quotation}}
\newenvironment{Presented}{\begin{quotation} \begin{center} 
             PRESENTED AT\end{center}\bigskip 
      \begin{center}\begin{large}}{\end{large}\end{center} \end{quotation}}

\begin{document}

\begin{titlepage}
 \pubblock
\vfill
\Title{Measurements of $W$ boson production in association with heavy flavour at ATLAS}
\vfill
\Author{Francesco Giuli (on behalf of the ATLAS Collaboration\footnote{Copyright 2023 CERN for the benefit of the ATLAS Collaboration. Reproduction of this article or parts of it is allowed as specified in the CC-BY-4.0 license.})}
\Address{CERN, EP Department, CH-1211 Geneva 23, Switzerland}
\vfill
\begin{Abstract}
The production of $W/Z$ bosons in association with heavy flavour jets or hadrons at the LHC is sensitive to the flavour content of the proton and provides an important test of perturbative QCD. The production of a $W$ boson in association with $D^{+}$ and $D^{*+}$ mesons will be discussed. This precision measurement provides information about the strange content of the proton. Measurements are compared to the state-of-the art next-to-next-to-leading order theoretical calculations.  
\end{Abstract}
\vfill
\begin{Presented}
DIS2023: XXX International Workshop on Deep-Inelastic Scattering and
Related Subjects, \\
Michigan State University, USA, 27-31 March 2023 \\
     \includegraphics[width=9cm]{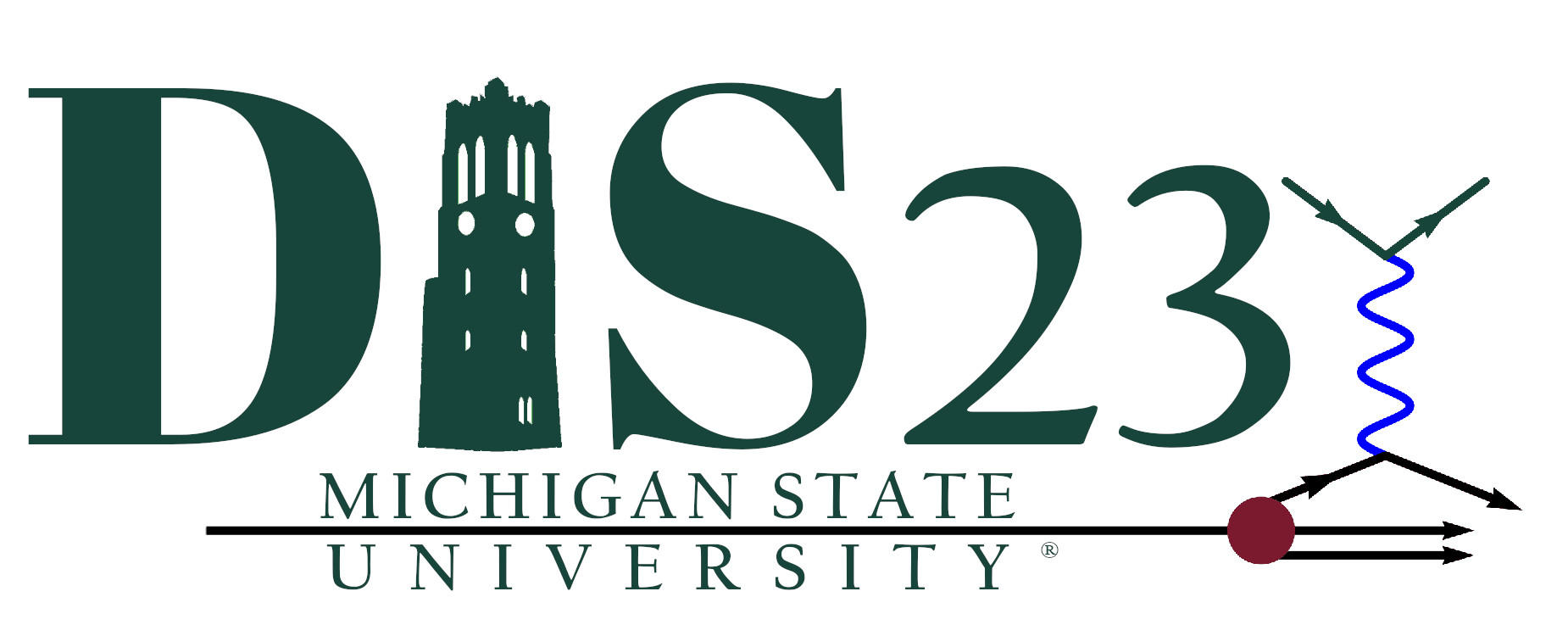}
\end{Presented}
\vfill
\end{titlepage}

\section{$W$ + charmed hadron production}
An accurate determination of the strange quark distribution functions (PDFs) of the proton is key to carrying out precision phenomenology at current and future colliders, specifically for measuring fundamental parameters of the Standard Model (SM), such as the mass of the $W$ boson. In perturbative quantum chromodynamics (pQCD), the production of a $W$ boson in association with a single charm quark occurs through the scattering of a gluon and a down-type quark, so it has a direct sensitivity to strange content of the proton.\\
This analysis presents a measurement of $W$ boson production in association with a $D^{(*)}$ meson~\cite{ATLAS:2023ibp},  using 140 fb$^{-1}$ $\sqrt{s}$ = 13~TeV proton–proton ($pp$) collision data recorded by the ATLAS detector~\cite{ATLAS:2008xda} at the LHC.  The following decay modes (and their charge conjugates) are used to study production of charmed hadrons: $D^{+}\rightarrow K^{-}\pi^{+}\pi^{+}$ and $D^{*+}\rightarrow D^{0}\pi^{+}\rightarrow (K^{-}\pi^{+})\pi^{+}$.  In $W+c$ production, at leading order (LO) the $W$ boson and charm quark always have opposite-sign electric charges, i.e. either $W^{+}+\bar{c}$ or $W^{-}+c$.  So inclusive and fiducial production cross-sections are extracted as the difference between the number of opposite-sign (OS) and same-sign (SS) events, since many of the backgrounds (i.e. $b$-hadron production from $t\bar{t}$ events and heavy-flavour (HF) pair production) are charge symmetric.  

\section{Event selection}
Events with leptons ($\ell = e$ or $\mu$) having a transverse momentum $p_{\mathrm{T}}^{\ell} >$ 30~GeV and pseudorapidity $|\eta(\ell)| <$ 2.5 are selected. Moreover, requirements for $D^{(*)}$ meson are applied, namely $p_{\mathrm{T}}(D^{(*)}) >$ 8 GeV and $|\eta(D^{(*)})| <$ 2.2. Furthermore,  in order to enhance the $W$ boson signal purity and reduce the multijet background, additional requirements are imposed: $E_{\mathrm{T}}^{\mathrm{miss}} > $ 30~GeV and $m_{\mathrm{T}} >$ 60~GeV, where $m_{\mathrm{T}}$ represents the $W$ boson transverse mass, defined as $\sqrt{2p_{\mathrm{T}}^{\ell}E_{\mathrm{T}}^{\mathrm{miss}}(1-\cos(\Delta\phi))}$, being $\Delta\phi$ the azimuthal separation between the lepton and the missing transverse energy.

\section{Signal and background modelling}
Monte Carlo (MC) $W+D^{(*)}$ events are categorised according to the origin of the tracks used to reconstruct the $D^{(*)}$ meson candidate, and this categorisation is based on MC truth information, namely:
\begin{itemize}
\item $W+D^{(*)}$ signal: if all tracks originate from the $D^{+}$ or $D^{*}$ hadrons and are assigned to the correct particle species $K^{\mp}\pi^{\pm}\pi^{\pm}$,
\item $W+c^{\mathrm{match}}$: if all the tracks originate either from a different hadron species or from a differed decay mode of a signal charmed meson,
\item $W+c^{\mathrm{mismatch}}$: if at least one but not all tracks belong to a single charmed hadron,
\item $W+\mathrm{jets}$: if none of the tracks are matched to a particle originating from a charmed particle.
\end{itemize}
While the $W+c^{\mathrm{match}}$ is modelled using \texttt{aMC@NLO+PY8} ~\cite{Alwall:2014hca, Sjostrand:2014zea} because the \texttt{EVTGEN}~\cite{Lange:2001uf} decays tables and models provide a better description of the $D$ meson decay rates and kinematics, the $W+c^{\mathrm{mismatch}}$ and $W+\mathrm{jets}$ backgrounds are modelled using \texttt{SHERPA}~2.2.11~\cite{Sherpa:2019gpd}, given its higher accuracy at Matrix-Element (ME) level.\\
Multijet backgrounds arise if one or more constituents of a jet are misidentified as a prompt lepton. This background rate is determined using the data-driven Matrix Method~\cite{ATLAS:2022swp},  since MC-based predictions for the normalization and composition of these backgrounds suffer from large uncertainties.\\
Finally, other background sources, such as processes containing top quarks ($t\bar{t}$, single-$t$ and $t\bar{t}X$) or from $Z+\mathrm{jets}$ and diboson, are modelled using MC simulation.
\begin{figure}[t!]
\begin{center}
\includegraphics[width=0.443\textwidth]{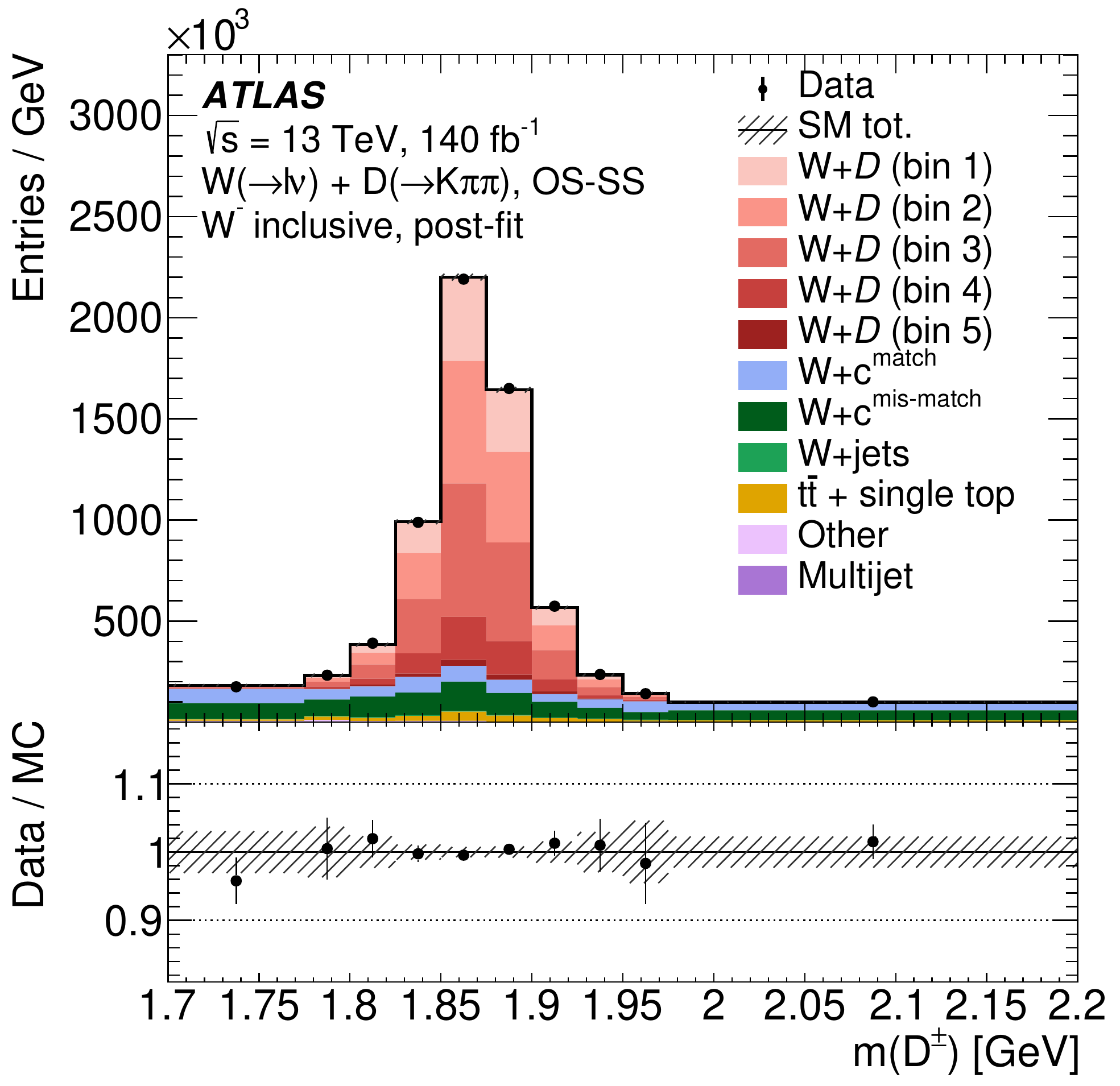}
\includegraphics[width=0.443\textwidth]{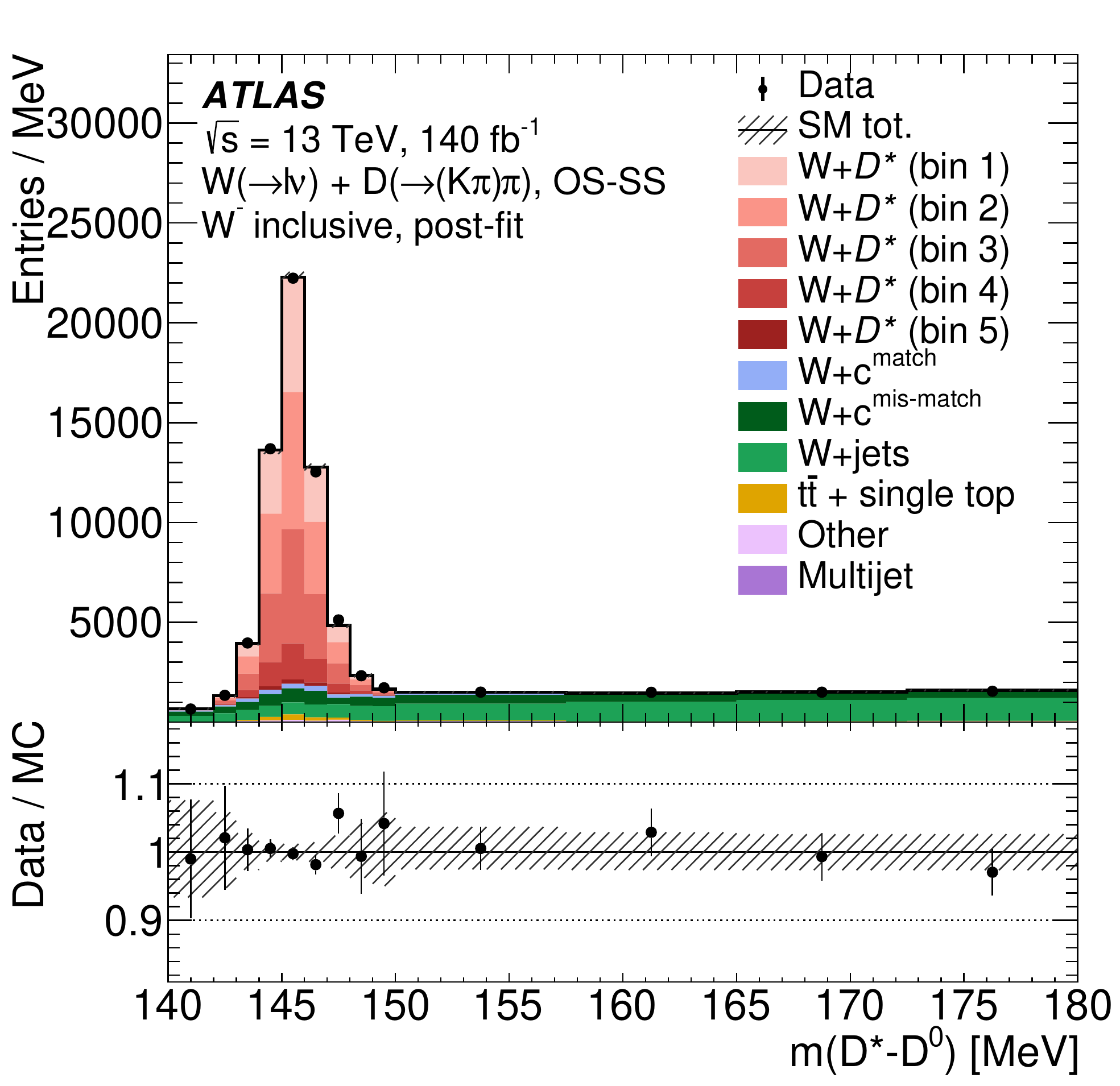}
\end{center}
\caption{Post-fit OS–SS $W+D^{(*)}$ signal and background predictions compared with data. The ``SM Tot.'' line represents the sum of all signal and background samples. In the panel, the data/MC ratio is shown, together with the corresponding hatched band showing the full post-fit systematic uncertainty. These plots are taken from Ref.~\cite{ATLAS:2023ibp}.} 
\label{fig:WD_kin}
\end{figure}

\section{Cross-section determination}
\begin{figure}[t!]
\begin{center}
\includegraphics[width=0.61\textwidth]{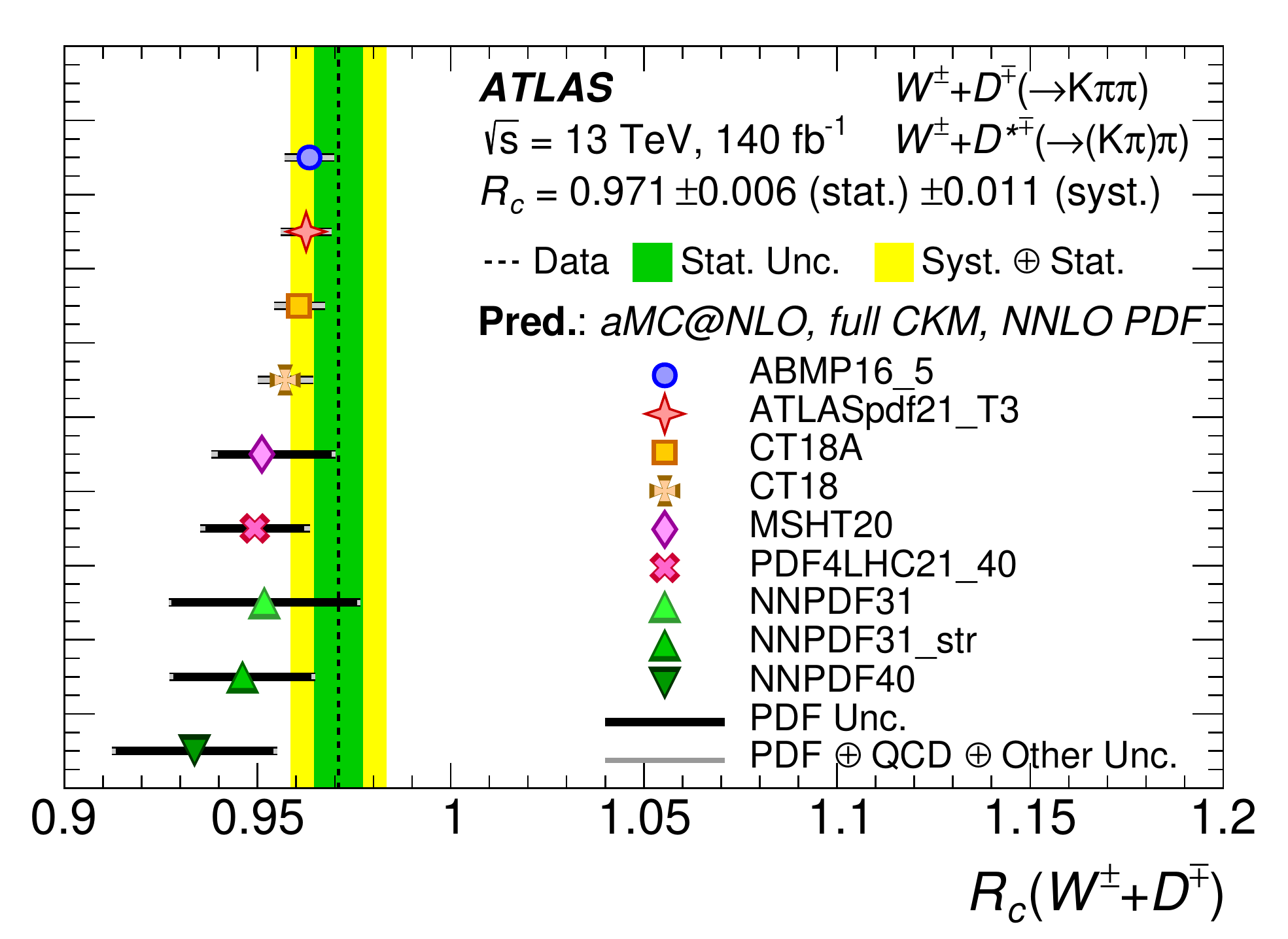}
\end{center}
\caption{Measured fiducial cross-section ratio, $R_{c}(W^{\pm}+D^{\mp})$, compared with different PDF predictions.  The predictions are based NLO calculations performed using \texttt{aMC@NLO} and a full CKM matrix with the following PDF sets: ABMP16{\_}5~\cite{Alekhin:2017kpj}, ATLASpdf21{\_}T3~\cite{ATLAS:2021vod}, CT18A, CT18~\cite{Hou:2019efy}, MSHT20~\cite{Bailey:2020ooq}, PDF4LHC21{\_}40~\cite{PDF4LHCWorkingGroup:2022cjn}, NNPDF31~\cite{NNPDF:2017mvq}, NNPDF31{\_}str~\cite{Faura:2020oom} and NNPDF40~\cite{NNPDF:2021njg}. All the PDF uncertainties are shown at 68\% CL. This plot is taken from Ref.~\cite{ATLAS:2023ibp}.} 
\label{fig:Rc}
\end{figure}
In order to measure the observables from the data with corresponding uncertainties,  a statistical fitting procedure based on the standard profile-likelihood formalism~\cite{Moneta:2010pm,Verkerke:2003ir} is used.  The extracted variables are: absolute fiducial cross sections $\sigma_{\mathrm{fid}}^{\mathrm{OS-SS}}(W^{-}+D^{(*)})$ and $\sigma_{\mathrm{fid}}^{\mathrm{OS-SS}}(W^{+}+D^{(*)})$, the cross section ratio $R_{c}^{\pm} = \sigma_{\mathrm{fid}}^{\mathrm{OS-SS}}(W^{+}+D^{(*)})/\sigma_{\mathrm{fid}}^{\mathrm{OS-SS}}(W^{-}+D^{(*)})$ and the differential cross-sections for OS-SS $W^{+}+D^{(*)}$ and $W^{-}+D^{(*)}$.  In situ systematic uncertainties constraints (using information from the data in the mass peak sidebands and control regions) and the estimation of background normalisation are enabled by the likelihood fit.\\
A binned likelihood function, $\mathcal{L}(\vec{\sigma},\vec{\theta})$, is constructed as the product of Poisson probability terms for each bin of the input mass distribution. In each bin of either $p_{\mathrm{T}}(D^{(*)})$ or $|\eta(\ell)|$, the product over the mass bins is performed, and $m(D^{+})$ ($m(D^{*+}-D^{0})$), the reconstructed invariant mass of the $D^{+}$ meson (the mass difference between the reconstructed invariant mass of the $D^{+}$ and $D^{0}$ mesons) is used as input in the $D^{+}$ ($D^{*}$) channel.\\
Within the likelihood fit, in order to perform the OS-SS subtraction, we developed a fitting procedure exploiting the charge correlation between the $W$ boson and the $D^{(*)}$ meson. Instead of using distributions after the OS-SS subtraction, both the OS and SS regions enter the likelihood function, with a common floating component added in both regions.

\section{Results and comparison with theoretical predictions}
\begin{figure}[t!]
\begin{center}
\includegraphics[width=0.443\textwidth]{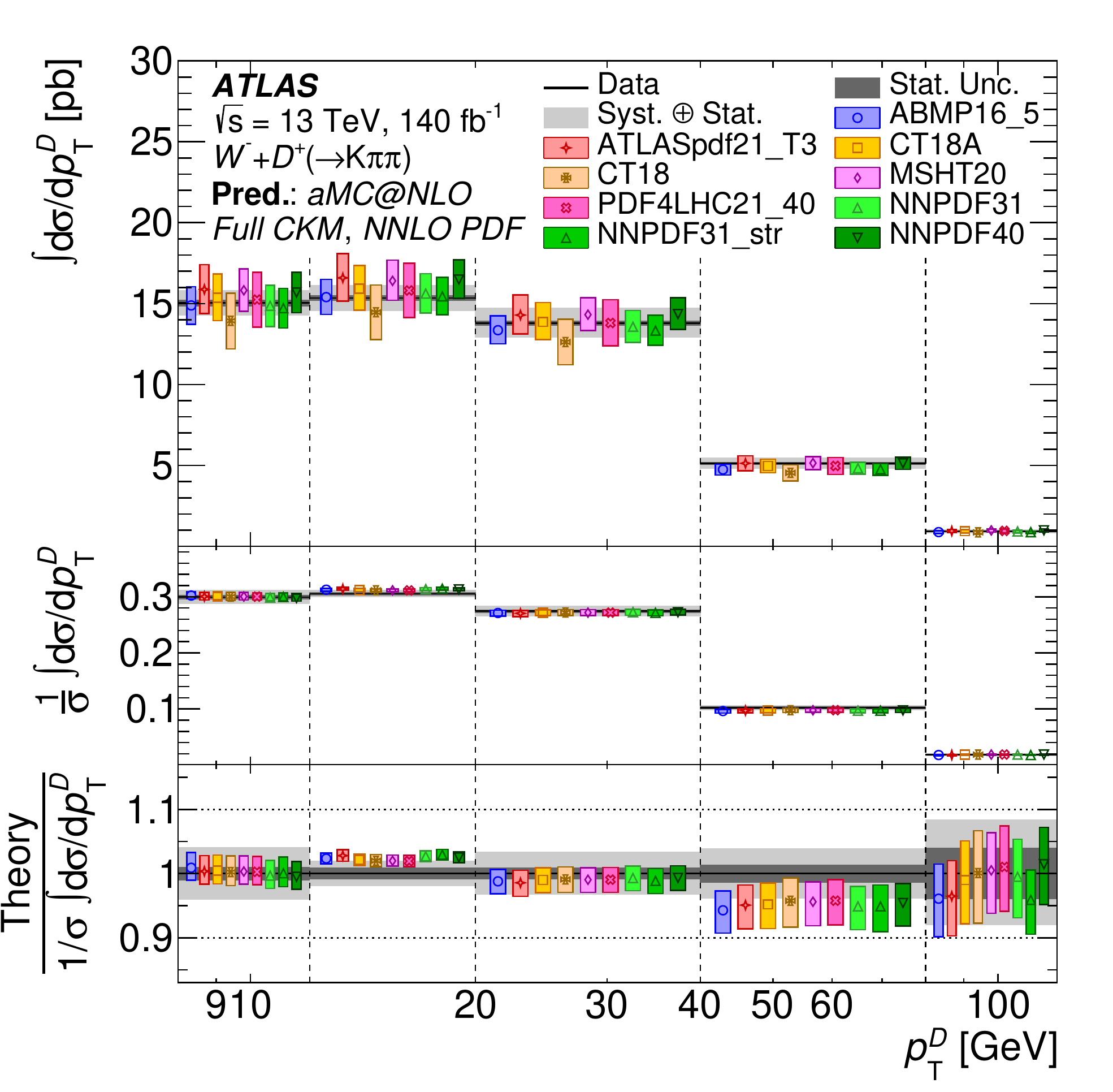}
\includegraphics[width=0.443\textwidth]{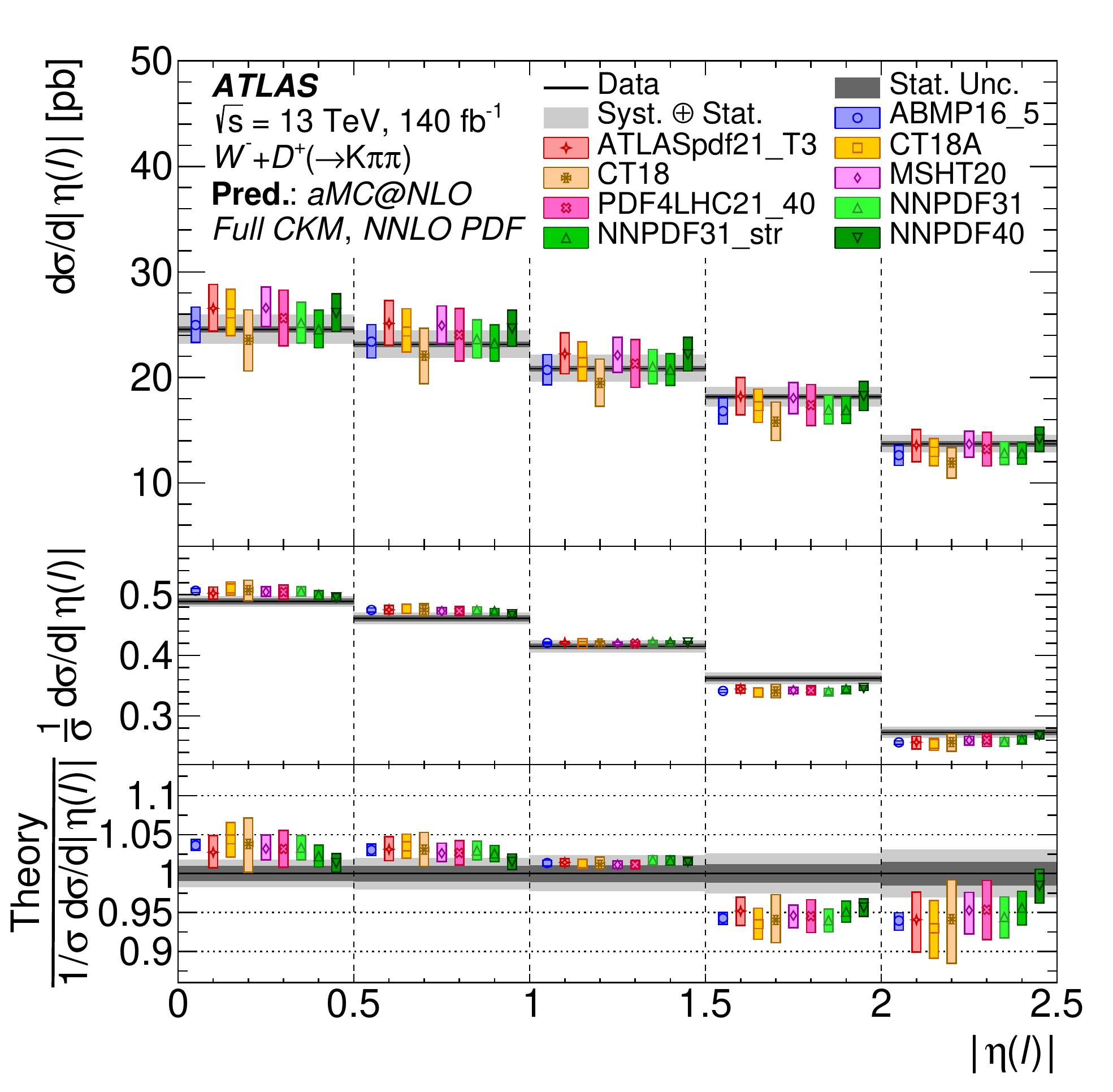}
\end{center}
\caption{Measured differential fiducial cross-section times the single-lepton-flavour $W$ branching ratio compared with different NNLO PDF predictions in the $D^{+}$ channel.  The displayed cross sections in $p_{\mathrm{T}}(D^{+})$ plots are integrated over each differential bin. The PDF predictions are based on NLO calculations performed using aMC@NLO and a full CKM matrix, using the same NNLO PDF set as in Figure~\ref{fig:Rc}. These plots are taken from Ref.~\cite{ATLAS:2023ibp}.} 
\label{fig:xsec_WD}
\end{figure}
Figure~\ref{fig:WD_kin} shows post-fit data/MC comparison plots for the $D^{+}$ and $D^{*+}$ channels for the $W^{-}$ channel. Most of the data points are within the resulting 1$\sigma$ uncertainty band, and similar agreement has been found in the $W^{+}$ channel as well.\\
Theoretical predictions of the $W+D^{(*)}$ cross sections are obtained using \texttt{aMC@NLO+PY8}. These predictions are based on next-to-leading order (NLO) QCD calculations performed using \texttt{aMC@NLO} and a full CKM matrix and they are computed for several next-to-next-to-leading order (NNLO) PDF set. \\
The cross-section ratio, $R_{c}(W^{\pm}+D^{\mp})$, is shown for the combined $D^{+}$ and $D^{*+}$ channels in Figure~\ref{fig:Rc}. Those sets that impose the restriction that the strange-sea be symmetric ($s=\bar{s}$), such as CT18 and ATLASpdf21, predict $R_{c}(W^{\pm}+D^{\mp})$ with higher precision while PDF fits that allow the $s$ and $\bar{s}$ distributions to differ, such as NNPDF or MSHT, have larger uncertainties. These measurements are consistent with the predictions obtained with PDF sets that impose a symmetric $s=\bar{s}$ sea, suggesting that any $s=\bar{s}$ asymmetry is small in the Bjorken-$x$ region probed by this measurement.\\
Figure~\ref{fig:xsec_WD} shows the differential cross sections, together with predictions obtained with the same NNLO PDF sets used in Figure~\ref{fig:Rc}. Consistency in the patterns observed in the $D^{+}$ and $D^{*+}$ channels has been found, for both the differential $|\eta(\ell)|$ and $p_{\mathrm{T}}(D^{*})$ cross-sections. It can be seen how the variations in shape of the two distributions depend only mildly on the PDF choice. Smaller systematic uncertainties have been found for $|\eta(\ell)|$ than $p_{\mathrm{T}}(D^{*})$, mostly because the secondary vertex reconstruction is independent of $|\eta(\ell)|$. Moreover, measurements of the cross-section as a function of $p_{\mathrm{T}}(D^{*})$ represents an important test of the quality of MC modelling of the production of heavy quarks, but they do not provide incisive constraints on PDFs. However, the systematic uncertainties for $|\eta(\ell)|$ are small and highly correlated among bins, providing good sensitivity to PDF variations. The significance of the discrepancy is reduced if the PDF uncertainties are considered, and this suggests that including these measurements in a global PDF fit would provide useful constraints on the allowed PDF variations.


\end{document}